\documentclass[12pt]{article}

\usepackage{graphicx}
\usepackage{amssymb}
\usepackage{amsmath}

\newcommand{\ket}[1]{|#1\rangle}

\newcommand {\micro}[1]{$\mu$#1}

\usepackage{scicite}

\usepackage{times}

\newenvironment{sciabstract}{%
\begin{quote} \bf}
{\end{quote}}

\newcounter{lastnote}
\newenvironment{scilastnote}{%
\setcounter{lastnote}{\value{enumiv}}%
\addtocounter{lastnote}{+1}%
\begin{list}%
{\arabic{lastnote}.}
{\setlength{\leftmargin}{.22in}}
{\setlength{\labelsep}{.5em}}}
{\end{list}}

\title{A quantum many-body spin system in an optical lattice clock}

\author
{M.~J.~Martin$^1$, M.~Bishof$^1$, M.~D.~Swallows$^1$, X.~Zhang$^1$,  C.~Benko$^1$,\\J.~von-Stecher$^1$,  A.~V.~Gorshkov$^2$, A. M.~Rey$^1$, and Jun~Ye$^{1}$\\
\normalsize{$^{1}$ JILA, National Institute of Standards and Technology and University of Colorado,} \\
\normalsize{and Department of Physics, University of Colorado, Boulder, CO 80309, USA.}\\
\normalsize{$^{2}$Institute for Quantum Information and Matter,} \\
\normalsize{California Institute of Technology, Pasadena, CA 91125, USA.}
}

\date{}

\begin{document} 


\baselineskip12pt


\maketitle


\begin{sciabstract}
Strongly interacting quantum many-body systems are fundamentally compelling and ubiquitous in science. However, their complexity generally prevents exact solutions of their dynamics. Precisely engineered ultracold atomic gases are emerging as a powerful tool to unravel these challenging physical problems. Here we present a new laboratory for the study of many-body effects: strongly interacting two-level systems formed by the clock states in ${}^{87}$Sr, which are used to realize a neutral atom optical clock that performs at the highest level of optical-atomic coherence and with precision near the limit set by quantum fluctuations. Our measurements of the collective spin evolution reveal signatures of many-body dynamics, including beyond-mean-field effects. We derive a many-body Hamiltonian that describes the experimental observation of severely distorted lineshapes, atomic spin coherence decay, density-dependent frequency shifts, and correlated quantum spin noise. These investigations open the door to exploring quantum many-body effects and entanglement in quantum systems with optical energy splittings, using highly coherent and precisely controlled optical lattice clocks.
\end{sciabstract}


\section{Introduction}

Strongly correlated quantum many-body systems have become a major focus of modern science. Researchers are using quantum-degenerate atomic gases\cite{Greiner:2002es,Bloch:2008gl, Lin:2009dr, Jo:2009jj, Will:2010dm, Simon:2011ep}, ultracold polar molecules\cite{Ni:2008wo, deMiranda:2011gd, Chotia2012}, and ensembles of trapped ions\cite{Kim:2010ib,Britton:2012gp} to realize novel quantum phases of matter and simulate complex condensed matter systems. In particular, the enhanced SU(\emph{N}) symmetry in the nuclear spin degrees of freedom of fermionic alkaline earth atoms may allow implementation of unconventional frustrated quantum magnetic models\cite{Gorshkov:2010hw,Wu2003,cazalilla2009,Hermele2009}. Although degenerate gases of alkaline earth(-like) atoms have been achieved\cite{Taie:2010hk,DeSalvo2010,Hara2011,Tey2010}, reaching the extremely low levels of entropy required to observe these novel magnetic phases is currently not feasible in ultracold atom experiments.

Meanwhile, optical atomic clocks employing fermionic alkaline earth atoms have matured considerably. The most stable of these clocks now operates near the quantum noise limit\cite{Nicholson:2012tl}, with an accuracy surpassing that of the Cs standard\cite{Ludlow:2008fd}. With atom-light coherence times reaching several seconds, even very weak interactions (\textit{e.g.}, fractional energy level shifts of order $\geq 1\times 10^{-16}$) can dominate the dynamics of these systems. In fact, atomic interactions in optical lattice clocks were first studied in the context of density-dependent frequency shifts that contribute to the clocks' systematic uncertainty\cite{Campbell:2009gaa,Lemke:2011ur, Swallows:2011er}. When frequency shifts were first observed in a ${}^{87}$Sr clock, they were attributed to $s$-wave collisions allowed by inhomogeneous excitation\cite{Campbell:2009gaa,Rey:2009eg,Gibble2009,Yu2010}, using a mean-field treatment and under the assumption that $p$-wave interactions were suppressed due to the $\sim$~1~\micro K sample temperature. More recently, in an optical clock based on ${}^{171}$Yb atoms at $\sim10$~\micro K, $p$-wave interactions were reported to lead to two-body losses and density shifts\cite{Lemke:2011ur,Ludlow:2011db}. At the same time, even at $\sim$~1~$\mu$K, inelastic $p$-wave losses were observed in the ${}^{87}$Sr system\cite{Bishof_PRA}. Although the importance of many-body interactions in these clocks has been recognized theoretically\cite{Rey:2009eg, Gibble2009,Yu2010}, to date no measurements have revealed their many-body nature.

In this Article, we report the first conclusive observation of beyond-mean-field many-body correlations in a high-density ${}^{87}$Sr optical clock in a one-dimensional (1D) optical lattice. We perform measurements near the standard quantum limit with ensembles of $\gtrsim 1000$ atoms. We show that non-classical correlations emerge as a consequence of collective elastic and inelastic $p$-wave interactions, which manifest themselves in the decay of Ramsey fringe contrast and as quadrature-dependent quantum noise in the effective spin degree of freedom encoded in the ${}^{1}\mathrm{S}_{0}$ and $ {}^{3}\mathrm{P}_{0}$ clock states. These effects cannot be captured by a conventional mean-field treatment of the atomic interactions. Instead, the quantum dynamics are precisely described by a many-body master equation that explicitly treats quantum fluctuations in the presence of two-body losses. In a prior experiment\cite{Swallows:2011er}, a strongly interacting regime (\textit{i.e.}, where atom-laser and atom-atom interactions are energetically comparable) was reached by tightly confining the atoms in a 2D optical lattice, at the expense of reducing the occupancy to one or two atoms per site.  Here we probe a strongly interacting system with an average of 20 atoms per disk-shaped 1D-lattice site, and we develop a detailed understanding of the complex many-body quantum dynamics. The role of $s$-wave collisions is suppressed by operating in the strongly interacting regime with highly homogeneous atom-laser coupling, making $p$-wave interactions, which operate collectively, dominant. We further show that the many-body dynamical evolution is well-described by a generalized infinite-range Ising model in an effective external magnetic field.

The experimental observation of such quantum magnetic behavior at $\mu$K temperatures is made possible because the motional degrees of freedom are effectively frozen during the clock interrogation. Only the internal electronic degrees of freedom (pseudo-spin) are relevant, and these can be initialized in a pure state. This Hamiltonian links the spin dynamics of interacting thermal fermions at $\mu$K temperatures to those of two-mode Bose-Einstein condensates (BEC) and it has been shown both theoretically\cite{Sorensen:2001jn,Kitagawa:1993vza,Wineland:1992bd,Steel:1998bw} and experimentally\cite{Esteve:2008ij, Gross:2010jn, Lucke2011} to give rise to non-trivial many-body correlations and quantum noise-squeezed states. It is also relevant in trapped-ion quantum simulation experiments\cite{Porras2004,Kim:2010ib,Britton:2012gp}.  The emergence of many-body correlations in optical lattice clocks subject to inelastic two-body losses opens up new scientific opportunities for the use of ultra-precise clocks as powerful tools for exploring strongly correlated open quantum systems\cite{Foss-Feig2012}.

\section{Many-body \textit{p}-wave model}

\begin{figure}
\centering
\includegraphics[width = 1 \columnwidth ]{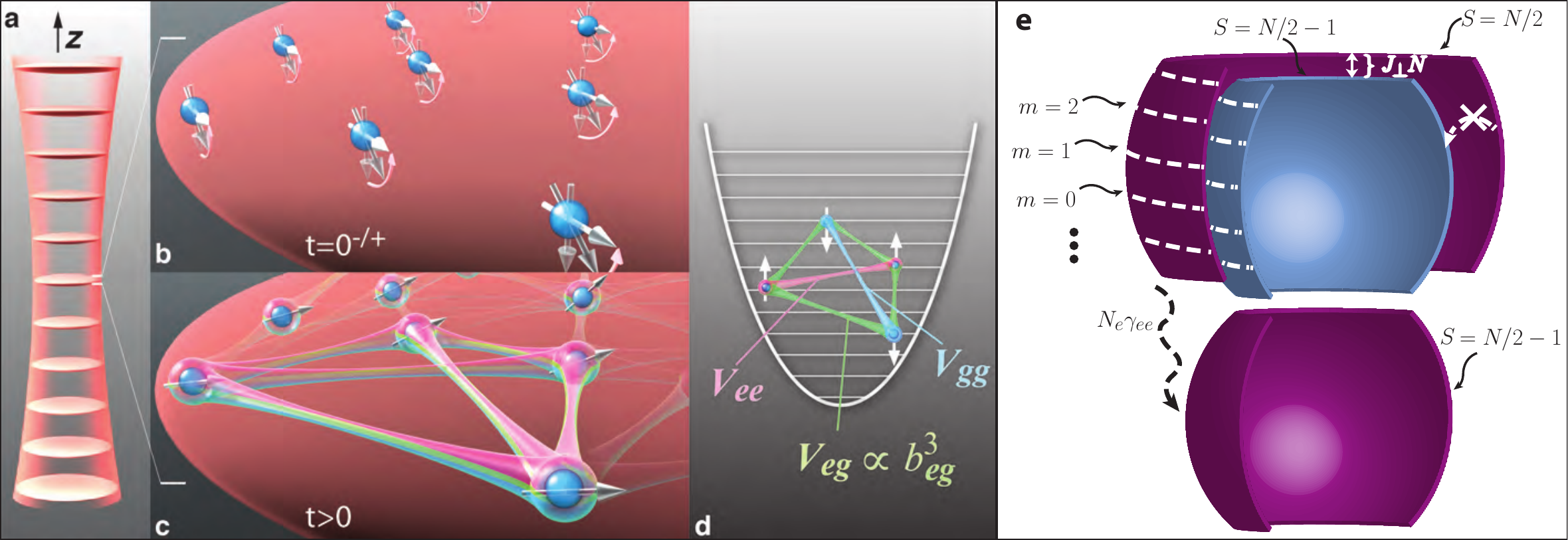}
\caption{\label{Fig1}
\textbf{Schematic diagram of the interacting many-body system. (a)} Approximately 100 lattice sites are significantly occupied during the experiment. The average lattice occupancy is 20 atoms. \textbf{(b)} At $t=0^{-}$, the particles are initialized in the ground state. After a short pulse, the atoms can be prepared in a product state of arbitrary single particle states, $\left(\alpha \ket{g} + \beta \ket{e} \right)^{{} \otimes N}$ at $t = 0^{+}$.  \textbf{(c)} The product state experiences the all-to-all many-body interaction for $t>0$.  \textbf{(d)} Interactions between particles are parametrized by the mode- and spin-dependent interaction parameters, $V_{gg} \propto b_{gg}^3$, $V_{eg} \propto b_{eg}^3$, and $V_{ee} \propto b_{ee}^3$ (see SOM for definition). \textbf{(e)} The many-body Hamiltonian has eigenstates comprised of maximally symmetric superpositions (Dicke states, for which $S =N/2$) of electronic ground and excited state superpositions, depicted as purple shells. Because of the slight inhomogeneities in the coupling strengths, the maximally symmetric manifold is coupled to the next lowest manifold with $S=N/2-1$, depicted as a nested blue shell. However, this coupling is prevented by an energy gap present in the many body Hamiltonian as a result of the $J^{\perp}  \vec{S} \cdot \vec{S}$ term, which is depicted by an offset of the two manifolds in this figure. Two-body inelastic losses connect maximally symmetric manifolds of $S\rightarrow S-1$, and thus are not a strong decoherence mechanism.}
\end{figure}

We consider an optical lattice clock that employs the  ${}^1$S$_0 \to {}^3$P$_0$ (henceforth  $\ket{g} \rightarrow \ket{e}$) clock transition
in nuclear spin-polarized ${}^{87}$Sr.  The lattice clock system is illustrated in Fig.~\ref{Fig1}a; it comprises an array of quasi-2D trap sites loaded with atoms at $\mu$K temperatures. Fig.~\ref{Fig1}b depicts the preparation of the system including state initialization (SOM). The tight lattice confinement along the longitudinal direction $Z$ freezes the dynamics and the population distribution among the trap sites along $Z$. A single site populated with $N$ atoms is modeled as a slightly anharmonic 2D oscillator with radial (longitudinal) frequency  $\nu_{R} =450$~Hz ($\nu_{Z} =80$~kHz).  The motional degrees of freedom remain frozen to the leading order and the dynamics take place only in the electronic pseudo-spin (SOM).
Additionally, the atoms are always prepared in the same state, thus Fermi statistics guarantee that no two atoms within a given trap site occupy the same motional state (Fig.~\ref{Fig1}d).

We consider the dynamics within a single trap site with $N$ atoms and label the thermally populated harmonic oscillator modes as ${\bf n}_j=(n_{X j},n_{Yj})$ with $j \in\{1,2,\dots N\}$. Under these conditions, the single-site dynamics can be described by an effective spin model (SOM). For  a pair of colliding  atoms occupying two of the $N$ eigenmodes, the interactions can be expressed in terms of spin $1/2$ operators ${\hat S}^{x,y,z}_{{\bf n}_j}$ acting  in the $\left\{e,g\right\}$ basis as: $J^{\perp}_{{\bf n}_j,{\bf n}_j'}\vec{{S}}_{{\bf n}_j}\cdot \vec{{S}}_{{\bf n}_j'} +\chi_{{\bf n}_j,{\bf n}_j'}{\Hat S}^{z}_{{\bf n}_j}{\Hat S}^{z}_{{\bf n}_j'}+ C_{{\bf n}_j,{\bf n}_j'}{\Hat S}^{z}_{{\bf n}_j'}$ . The subscripts  ${\bf n}_j,{\bf n}_j'$ indicate that the spin coupling constants in principle depend on single-particle modes populated by the atoms. The thermal averages of the mode-dependent coupling constants $J^{\perp}_{{\bf n}_j,{\bf n}_j'}$, $\chi_{{\bf n}_j,{\bf n}_j'}$, and $C_{{\bf n}_j,{\bf n}_j'}$ are sharply peaked about their averages $J^{\perp}$, $\chi$, and $C$ with standard deviations $\Delta J^{\perp}$, $\Delta \chi$, and $\Delta C$, respectively. However,
due to the weak dependence of the matrix elements on the thermally populated modes, $J^{\perp}\gg \Delta \chi, \; \Delta C $, and thus the large energy gap created by $J^{\perp}$ suppresses transitions between manifolds with different total collective spin  $S$ (Fig.~\ref{Fig1}e), caused by the inhomogeneities $\Delta \chi$ and $\; \Delta C$.
Here, $S(S+1)$ is the eigenvalue of the collective operator  $\vec{S} \cdot \vec{S}$ and $\hat S^{\tau=x,y,z} = \sum_{j=1}^N {\hat S}^{\tau}_{\mathbf{n}_j}$. The dynamics of atoms prepared in the  $S=N/2$ manifold is thus characterized by the collective Hamiltonian in the rotating frame of the laser, given by
 \begin{align}
&\hat H^{\mathrm{eff}}/ \hbar =  - \delta \hat S^{z}  - \Omega \hat S^{x} +  \chi \left(\hat S^{z}\right)^{2}+ C \left(N-1\right) \hat S^{z} + \mathcal{O}\left(S^{z}\right)^{3},
\label{Hamiltonian}
\end{align} where constants of motion have been omitted. Here, $\delta = \omega_{L} -\omega_{a}$ is the laser detuning from the atomic resonance and $\Omega$ is the Rabi frequency. The terms $\chi$ and $C$ originate from elastic $p$-wave interactions and are proportional to the  $p$-wave scattering volumes, $b_{gg}^3$, $b_{eg}^3$ and $b_{ee}^3$ (Fig.~\ref{Fig1}d). The matrix elements are averaged over the set of populated modes. The role of $s$-wave interactions in the absence of excitation inhomogeneity is limited to an enlargement of $J^{\perp}$, thus enhancing the gap protection. In a 2D trap geometry, $\chi$ and $C$ are highly insensitive to the sample temperature $T$ for a fixed atom number, given that the linear growth of the $p$-wave interaction with $T$ is compensated by a corresponding $1/T$ decrease of the density. The term of order $\left(\hat{S}^{z}\right)^{3}$ arises when higher order corrections are used to account for virtual excitations of motional states (SOM).  Similar 
effects have been measured in bosonic systems\cite{Johnson:2009um}.  Equation~\ref{Hamiltonian} is the key theoretical result of this work and describes the collective spin dynamics (Fig.~\ref{Fig1}c).

In addition to the elastic interaction, the effect of inelastic $p$-wave two-body losses must be included. In contrast with ${}^{171}$Yb, where inelastic decay channels exist for both excited-excited and excited-ground collisions\cite{Lemke:2011ur}, for ${}^{87}$Sr only excited-excited decay is relevant\cite{Bishof_PRA}. The correct treatment of the interplay between elastic and inelastic interactions during the many-body dynamics is challenging due to the  computational complexity of numerical techniques for open quantum systems. We capture the full many-body dynamics by solving a master equation that is numerically tractable owing to the simplification allowed by the weak $\{\mathbf{n}_{j}\}$-dependence of the $p$-wave interaction matrix elements. This leads to a two-body decay that preserves the collective nature of the model to leading order (Fig.~\ref{Fig1}c), allowing the master equation to be solved efficiently for systems of up to 50 atoms (SOM).

\section{Density shift and linear response}

\begin{figure}
\centering
\includegraphics[width = .75 \columnwidth ]{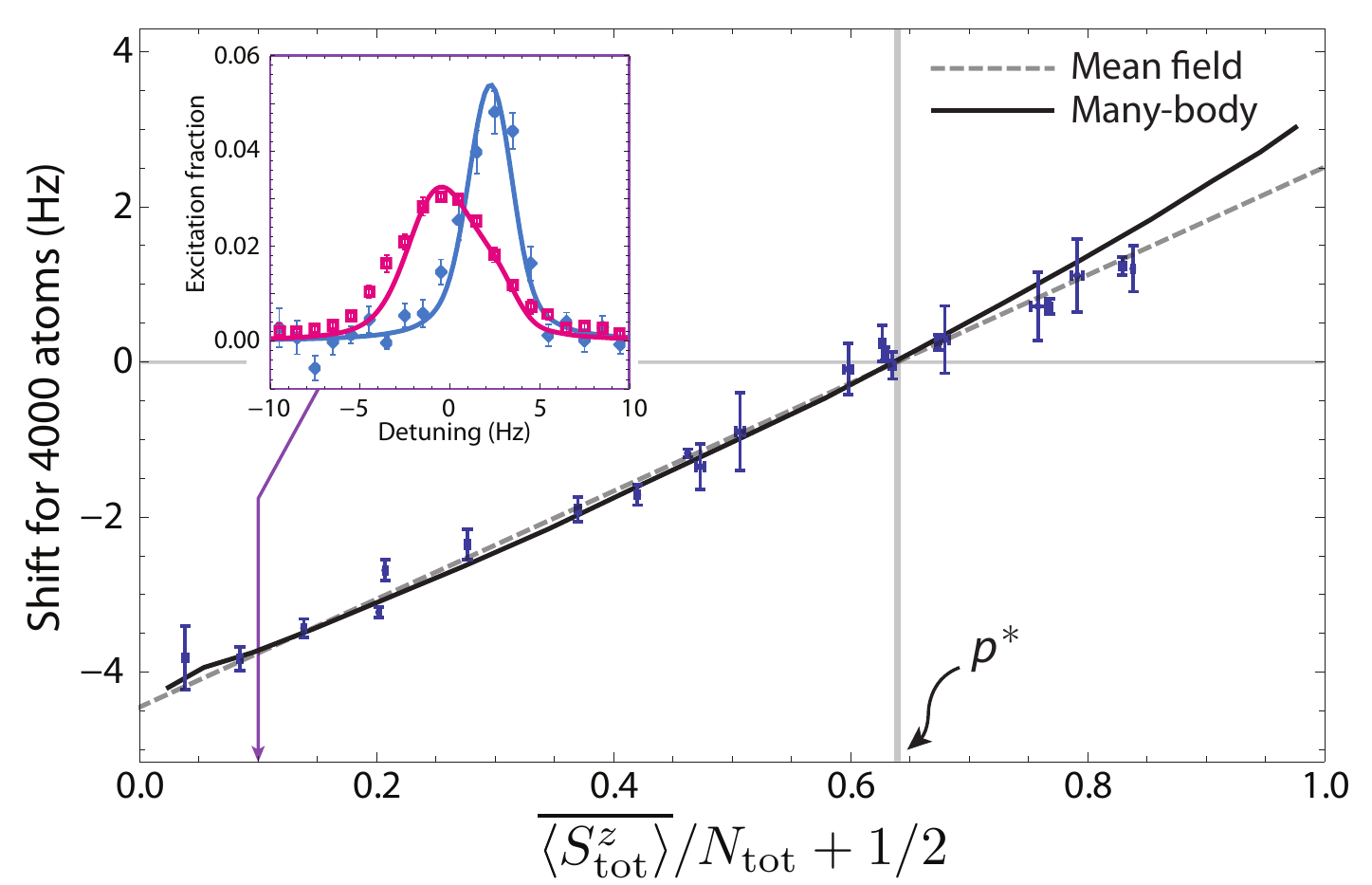}
\caption{\label{Fig2}
\textbf{Density shift in Ramsey spectroscopy fit with the full many-body solution.} Due to the perturbative nature of this measurement, the mean-field approximation to the many-body theory (dashed line) agrees well with the data. The exact many-body solution in the absence of losses (solid curve) agrees best with the data only for lower values of total average spin $\overline{\langle \hat{S}_{\mathrm{tot}}^{z}\rangle}/N_{\mathrm{tot}}$ due to the nonlinear $\left(S^{z}\right)^{3}$ term in the Hamiltonian. The zero crossing occurs at an average excitation fraction, given by $\overline{\langle \hat{S}_{\mathrm{tot}}^{z}\rangle}/N_{\mathrm{tot}} +1/2$, of $p^{\star}= 0.64(1)$. From the zero crossing and the measured slope we extract $\chi$ and $C$. At $\overline{\langle \hat{S}_{\mathrm{tot}}^{z}\rangle}/N_{\mathrm{tot}} \simeq \pm 1/2$, the dephasing due to population differences in sites is maximized, resulting in maximal Ramsey fringe decoherence (shown in Fig.~\ref{Fig4}). Near the zero-crossing, the Ramsey decoherence is dominated by intra-site effects due to the nonlinear term in the many-body Hamiltonian. \textbf{Inset:} linear response of the Rabi spectroscopic line shapes under weak excitation (pulse area $\theta_1 = 0.2 \pi$). The shift of the line center provides an independent verification of the model in the linear response regime.}
\end{figure}

Using a modified Ramsey spectroscopy sequence, we measure the density-dependent frequency shift of the clock transition to determine the relevant interaction parameters that characterize our spin Hamiltonian, Eqn.~\ref{Hamiltonian}. The initial pulse area $\theta_{1} = \Omega T_{R}$, chosen such that $0<\theta_{1}<\pi$,  controls the initial value of $\langle \hat S_{\mathrm{tot}}^{z}\rangle$. Here, $\hat S^{z}_{\mathrm{tot}}$ is the sum of $\hat S^{z}$ over the $\sim 100$ relevant sites, such that $-N_{\mathrm{tot}}/2 \leq \langle \hat S_{\mathrm{tot}}^{z}\rangle \leq N_{\mathrm{tot}}/2$, where $N_{\mathrm{tot}}$ is the total number of atoms loaded into the lattice. In the presence of two-body losses, $\langle \hat S_{\mathrm{tot}}^{z}\rangle$ is not constant, thus we use its time average, $\overline{\langle \hat S_{\mathrm{tot}}^{z}\rangle}$, to facilitate comparison with theory. Here, we extract $\overline{\langle \hat S_{\mathrm{tot}}^{z}\rangle}$ from independent measurements periodically inserted into the clock sequence. The duration of the dark time, $\tau_{\mathrm{dark}}$, is fixed at 80~ms and the final pulse area is set to $\pi/2$.  We measure the shift by modulating the density by a factor of $\sim$ 2 (Fig.~\ref{Fig2}).

A simple mean-field analysis of Eqn.~\ref{Hamiltonian} (neglecting cubic terms and losses)  treats interactions as an effective magnetic field along $Z$,  $B(N)=N C + 2 \chi \langle \hat S^z\rangle$, where $\langle \hat S^{z}\rangle =-(N/2) \cos\theta_{1}$. The mean-field density-dependent frequency shift $\Delta \nu(N)=B(N)/ \left(2\pi \right)$ scales linearly with the excitation fraction and agrees with experimental observations (Fig.~\ref{Fig2}).  Additionally, we fit an exact solution of Eqn.~\ref{Hamiltonian} to the data. Both fits are shown in Fig.~\ref{Fig2}. To compare with the experiment, we always perform an average over the atom number distribution across the lattice sites. From this measurement we extract $\chi = 2 \pi \times 0.20(4)$~Hz and $C = - 0.3 \chi$.
The linear behavior of the density-dependent frequency shift with Ramsey spectroscopy was also measured in the ${}^{171}$Yb lattice clock\cite{Lemke:2011ur,Ludlow:2011db}.

As a complementary measurement, we further probe the linear response dynamics of the system spectroscopically with a single weak pulse while varying $\delta$. The pulse duration is $T_{R} =500$~ms, and its area is $\Omega T_{R} \simeq 0.2 \pi$. At this low excitation fraction, two-body losses can be ignored, and the mean-field analysis yields a density shift of the clock resonance given by $\Delta \nu_{\mathrm{LRR}} \simeq (C-\chi) N$, which agrees with the mean-field expression of the Ramsey frequency shift at small $\theta_{1}$.  In the experiment, we observe a shift of 2.7~Hz for a modulation between $N_{\mathrm{High}}=3\times10^3$ and $N_{\mathrm{Low}}=1\times 10^3$ (Fig.~\ref{Fig2}, inset). We extract $\chi$ from a simple mean-field model and find agreement at the 15$\%$ level, consistent with the atomic number distribution uncertainty.
\label{ShiftSection}

\section{Rabi spectroscopy}
Having extracted the interaction parameters as described in Section~\ref{ShiftSection}, we turn our attention to regimes where strong interaction effects emerge by probing the system via Rabi spectroscopy. The atom-laser interaction energy, $\hbar \Omega$, is reduced to a level comparable to the many-body interaction energy. The pulse duration, $T_{R}$, becomes correspondingly long to allow for high spectroscopic resolution of the many-body excitation spectra. Typically we set $\Omega T_{R} = \pi$.  In terms of the spin Hamiltonian, the evolution now takes place in the presence of a transverse magnetic field-like term that does not commute with the Ising interactions, thus giving rise to non-trivial quantum behavior.  In a direct analogy to a double-well potential for BEC, $\Omega$ and $\delta$ play the role of tunneling and bias between the left and right wells, respectively; and their competition with non-linear interactions has been shown to lead to novel quantum dynamics including macroscopic quantum self-trapping\cite{Trombettoni2001,Albiez2005}.

\begin{figure}
\centering
\includegraphics[width=  1\columnwidth]{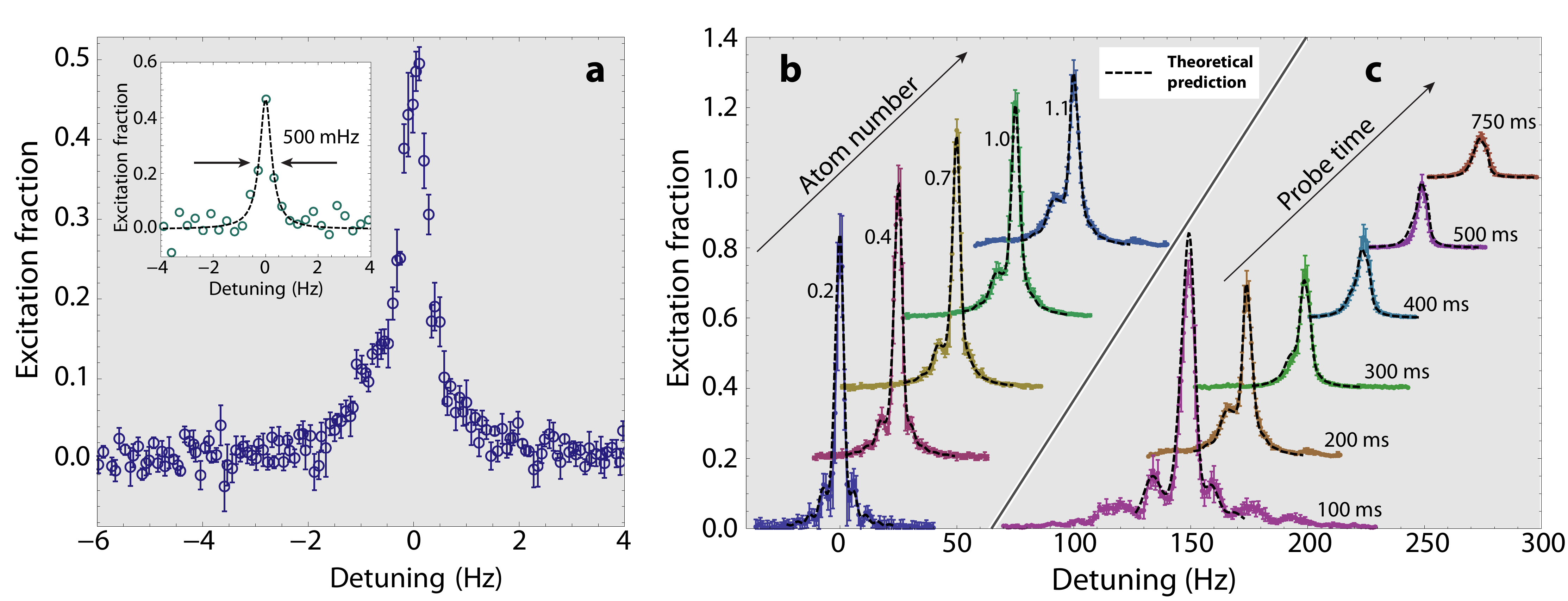}
\caption{\label{Fig3}
\textbf{Lineshape as a function of experimental parameters.} Each curve is a composite of multiple scans that have been centered atop one another. These data are subsequently binned with a binsize of 1~Hz. The curves are offset in both the vertical and horizontal directions for visual clarity. \textbf{(a)} Lineshape obtained at an operating density of $0.1 \rho_{0}$ with a 2~s clock probe, where $\rho_{0}$ is the density obtained for $5\times10^{3}$ atoms, and is given by $\rho_{0} =5(2)\times 10^{11}$~cm$^{-3}$. The line shape shows a clear distortion due to the density-dependent interaction at the sub-Hz energy scale. \textbf{(Inset)} Extremely low density scan ( $\rho=5 \times 10^{-2} \rho_{0}$) with a 3~s clock probe, demonstrating the ultimate frequency resolution of the system.  \textbf{(b)} Lineshape as a function of density (normalized by $\rho_{0}$), with a $\pi$-pulse time of 200~ms.  \textbf{(c)} Lineshape for $\rho \simeq \rho_{0}$ as a function of probe time.  In both \textbf{(b)} and \textbf{(c)}, theoretical curves, obtained with the mean-field treatment including loss, are shown as dashed black lines and agree well with the measurements.
}
\end{figure}

At the highest achievable densities in our experiment, we observe dramatic deviations in the Rabi lineshapes from ideal, single-particle lineshapes.
By varying either the atom number or $\Omega$ (with $T_{R} = \pi/\Omega$), we change the ratio of the many-body interaction energy to $\hbar \Omega$. The resulting lineshapes are summarized in Fig.~\ref{Fig3}. Our single-scan frequency resolution is $500$~mHz, and we are thus capable of resolving sub-Hz interaction features spectroscopically (Fig.~\ref{Fig3}a). At low atom number ($N_{\mathrm{tot}} \simeq 1\times10^{3}$) and with $T_{R}$ = 200~ms, we observe a nearly ideal Rabi lineshape that becomes severely distorted with increasing density (Fig.~\ref{Fig3}b). Similarly, as we increase $T_{R}$, keeping $\Omega T_{R} = \pi$, we begin to see the onset of an interaction blockade mechanism (macroscopic self-trapping in the BEC context\cite{Trombettoni2001}) in concert with the distortion (Fig.~\ref{Fig3}c). At the largest $T_R$ = 750~ms, we excite only 20\% of the atoms into $\ket{e}$ and the line is approximately five times broader than the non-interacting lineshape.

Using the interaction parameters extracted via the Ramsey density shift and the linear response measurements, we are able to theoretically reproduce all the observed lineshapes (theoretical curves shown in Fig.~\ref{Fig3}b--c) using entirely the mean-field treatment of Eqn.~\ref{Hamiltonian} in the presence of two-body losses. Here, a full many-body treatment of the master equation is in agreement with the mean-field predictions.

\section{Quantum fluctuations and beyond mean-field physics}

While the density-dependent frequency shift and lineshape measurements provide verification for the mean-field limit of the many-body model and are useful for extracting the interaction parameters in a simple and efficient way, they do not represent a sensitive probe of the full many-body dynamics. Ramsey fringe contrast, on the other hand, can undergo a periodic series of collapses and revivals,
which will reflect the quantized structure of the many-body spectrum arising from the granularity of the interacting atoms. Specifically, for a fixed $N$ and neglecting both the losses and the $(\hat S^z)^3$ term, the amplitude of the Ramsey fringe evolves as $Z^{N-1}$ with $Z^2\equiv 1-\sin^2(\theta_1 ) \sin^2 \left (\chi \tau_{\mathrm{dark}}\right)$.  At short times (\textit{i.e.}, $\sqrt{N} \chi \tau_{\mathrm{dark}} \ll1$), the contrast is approximately $1- \left(N-1\right) \chi^2 \sin^2(\theta_1 )\tau_{\mathrm{dark}}^2/2$. On the contrary, the mean-field model at the single-site level (with fixed $N$) predicts no decay of the Ramsey fringe contrast, as a magnetic field $B(N)$ can only lead to a pure precession of the collective Bloch vector. By taking the average over atom distributions among lattice sites and properly treating two-body loss during the Ramsey dark time, the mean-field model does show a decay of the contrast. However, this decay is associated mainly with dephasing arising from different precession rates exhibited by sites with different $N$. This effect is most relevant for low excitation fractions where $B(N)$ is large and the inelastic loss is suppressed.

We implement a Ramsey sequence to experimentally measure the fringe contrast. The Ramsey pulse durations are $<6$~ms, satisfying $\Omega \gg N \chi$, to suppress interaction effects during the pulses. We apply the final $\pi/2$ readout pulse with a variable relative optical phase of $0^{\circ}$ -- $360^{\circ}$. We record the fraction of excited atoms as a function of the readout phase, which constitutes the Ramsey fringe. The Ramsey fringe contrast is extracted in a manner that is insensitive to the frequency noise of the ultrastable clock laser (SOM).

\begin{figure}
\centering
\includegraphics[width =  1 \columnwidth ]{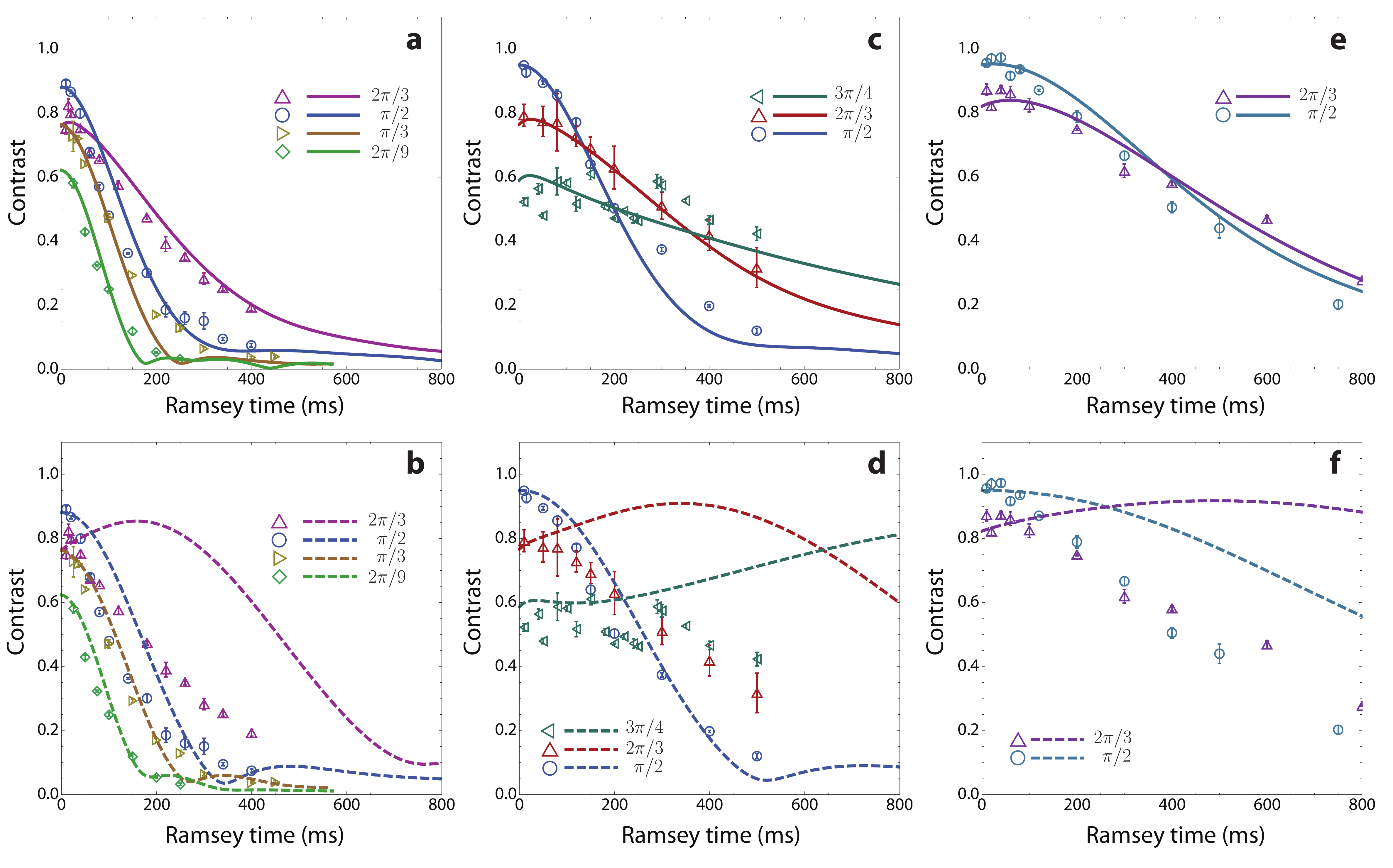}
\caption{\label{Fig4}
\textbf{Ramsey fringe contrast decay vs initial pulse area.}  The pulse area values for the data and corresponding theory are given in the legends of the plots. Error bars represent the statistical error of each contrast measurement, and thus do not account for systematic drifts that occur over the course of the experiment. The solid lines (top panels) are the many-body calculations, while the dashed lines (bottom panels) are using the mean-field approximation of the theory. The many-body model and the mean-field approximation agree in the limit of small initial pulse area  (\textit{i.e.}, Bloch vector polar angle), but disagree for pulse areas $\gtrsim \pi/2$. This is an important confirmation of the dominance of many-body effects in this parameter regime.  \textbf{(a, b)} $\nu_{Z}=80$~kHz, $\nu_{R}=450$~Hz, and $N_{\mathrm{tot}} = 4000$; \textbf{(c, d)} $\nu_{Z}=65$~kHz, $\nu_{R}=370$~Hz, and $N_{\mathrm{tot}} = 4000$; and \textbf{(e, f)} $\nu_{Z}=80$~kHz, $\nu_{R}=450$~Hz, and $N_{\mathrm{tot}} = 1000$.}
\end{figure}

We explore three distinct experimental conditions to rule out single-particle decoherence mechanisms and thoroughly test the model. The first condition represents the typical operating parameters of the lattice clock, with $N_{\mathrm{tot}} = 4\times 10^{3}$ and $\nu_{Z} = 80$~kHz. In the second case, we reduce the lattice intensity such that $\nu_{Z} =65$~kHz, which results in a reduction of the density by a factor of $\sim$1.8.  Finally, we maintain $\nu_{Z} = 80$~kHz but reduce the atom number to $N_{\mathrm{tot}} = 1\times 10^{3}$. Under all conditions, the full many-body density matrix model reproduces the experimental observations well (Figs.~\ref{Fig4}a, \ref{Fig4}c, and \ref{Fig4}e). The inclusion of the $(\hat{S}^z)^3$ correction improves the theory-experiment agreement, especially for pulse areas $\theta_1>\pi/2$  and for the high-density conditions (SOM). We also observe a striking breakdown of the mean-field model for $ \theta_1 \gtrsim \pi/2$ where many-body corrections are expected to be dominant.

The frequency shift, lineshape, and Ramsey fringe contrast are quantities that all depend on the first-order expectation values of the spin operators $\langle \hat{S}^{x,y,z} \rangle$. We now turn our attention to the distribution of quantum noise, which depends on the second-order moments of the spin operators, \textit{e.g.}, $\langle  \left(\hat S^{x}\right)^{2} \rangle-\langle \hat{S}^{x} \rangle ^{2}$, $\langle \hat{S}^{x} \hat{S}^{z} + \hat{S}^{z} \hat{S}^{x}\rangle-2\langle  \hat{S}^{x} \rangle \langle \hat{S}^{z} \rangle  $, etc. Given that the form of the Hamiltonian in Eqn.~\ref{Hamiltonian} is known to produce squeezed and entangled states\cite{Kitagawa:1993vza}, the distribution of the quantum noise becomes a compelling measurement to probe physics beyond the mean field. However, until this work, optical clocks have lacked the requisite laser precision to probe the quantum noise distribution. Recent advances in ultrastable lasers have permitted access to the quantum regime\cite{Nicholson:2012tl}.

\begin{figure}
\centering
\includegraphics[width = 0.65 \columnwidth]{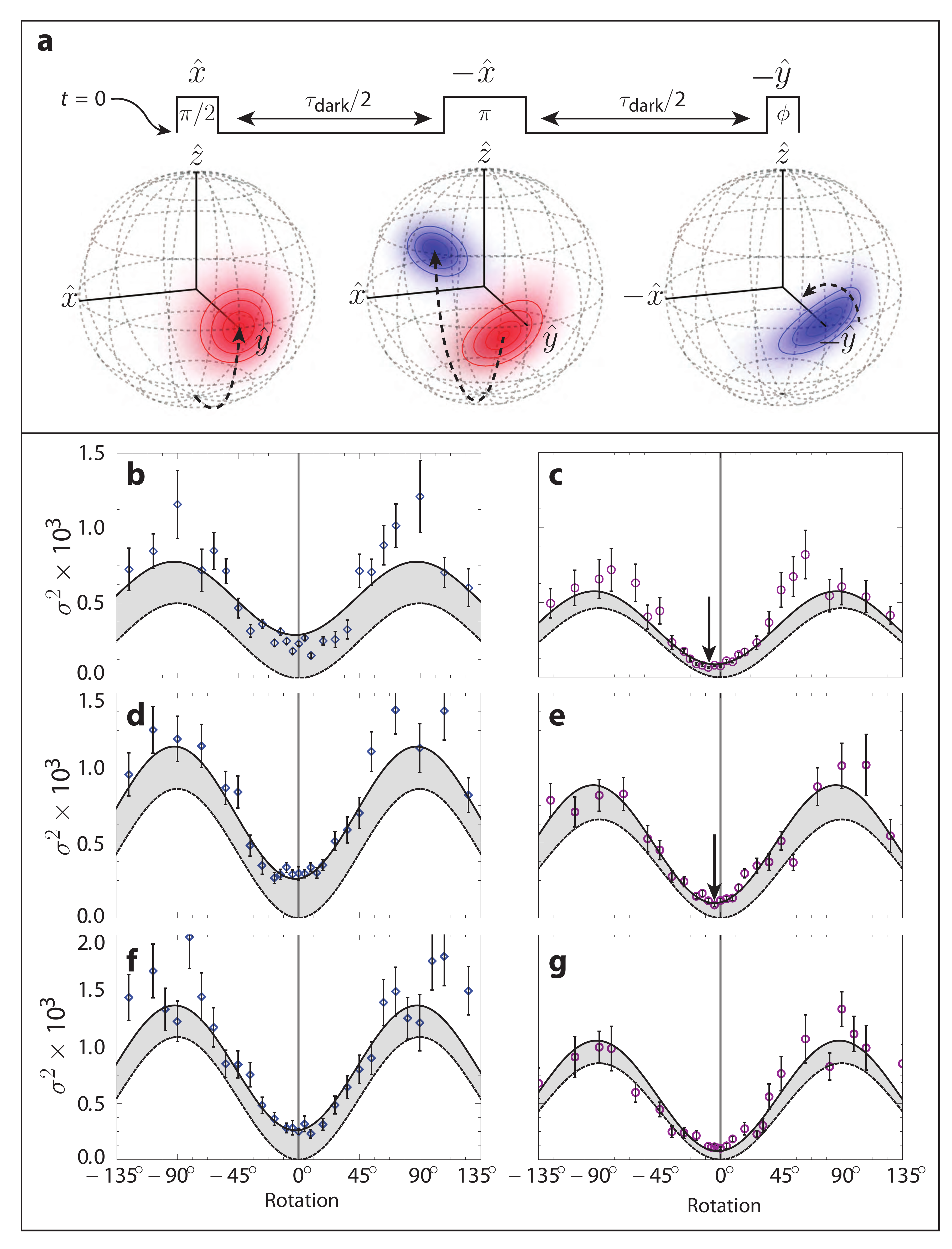}
\caption{\label{Fig5}
\textbf{Spin noise vs. quadrature.} \textbf{(a)} An initial pulse prepares a coherent state along $\hat{y}$, which then evolves for $\tau_{\mathrm{dark}}/2$. An echo pulse then rotates the many-body state $180^{\circ}$ about $-\hat{x}$. After and additional evolution time of $\tau_{\mathrm{dark}}/2$, a final pulse rotates the state about $-\hat{y}$ and the spin noise is measured. The many-body state depicted here represents the spin evolution of a 20 atom ensemble in a single trap site with $\tau_{\mathrm{dark}}$ = 40 ms. To remove spurious effects due to slow drifts in atom number, the data is processed as detailed in the SOM, which removes the potential bias.
For plots (b--g), the dashed line is the pure laser noise extracted from a fit to the data. The solid line is the laser noise plus the full many-body prediction of the spin noise. This full theory is simultaneously fit to both the low and high atom number curves to extract the laser noise for a given dark time. Vertical arrows indicate significantly phase-shifted minima in the experimentally measured spin noise, consistent with the predictions of the many-body theory.
Plots \textbf{(b, c)}: Spin noise for $N_{\mathrm{tot}}=1\times10^{3}$ and $N_{\mathrm{tot}}=4\times10^{3}$, respectively, with an interaction time of $\tau_{\mathrm{dark}}$ = 20~ms and a $\pi$-pulse time of 10~ms.  A clear phase shift is observed near the origin for the high atom number case. \textbf{(d, e)}:  Spin noise for $N_{\mathrm{tot}}=1\times10^{3}$ and $N_{\mathrm{tot}}=4\times10^{3}$, respectively, with $\tau_{\mathrm{dark}}$ = 40~ms and a $\pi$-pulse time of 20~ms.  In plot (d), a clear phase shift is also observed near the origin. \textbf{(f, g)}: Spin noise for $N_{\mathrm{tot}}=1\times10^{3}$ and $N_{\mathrm{tot}}=4\times10^{3}$, respectively, with $\tau_{\mathrm{dark}}$ = 60~ms and a $\pi$-pulse time of 20~ms.}
\end{figure}

To minimize single particle decoherence effects and dephasing due to the distribution of site occupancies, we add a spin-echo pulse to the Ramsey sequence.
As a result, the sensitivity to low-frequency laser noise is reduced at the expense of increased sensitivity to high-frequency laser noise. With atoms initialized in $\ket{g}$, we follow the pulse sequence shown in Fig.~\ref{Fig5} to manipulate and measure the spin noise of the many-body state.
For each value of the final rotation angle, representing a specific quadrature in which we measure the spin noise, we repeatedly record $\langle \hat{S}_{\mathrm{tot}}^{z}\rangle/N_{\mathrm{tot}}$ via measurements of the final atomic excitation fraction after the Ramsey sequence. From the data, we determine $\sigma^{2} \equiv  \langle \left(\hat S_{\mathrm{tot}}^{z} \right)^{2} \rangle/N_{\mathrm{tot}}^{2}  - \langle \hat{S}_{\mathrm{tot}}^{z} \rangle ^{2}/N_{\mathrm{tot}}^{2}$ by analyzing the pair variance for successive measurement of $\langle \hat{S}_{\mathrm{tot}}^{z}\rangle/N_{\mathrm{tot}}$. We note that the quantum limit of $\sigma^{2}$ is important for defining the ultimate stability of lattice clocks\cite{Nicholson:2012tl}.  For an ideal coherent spin state of the entire ensemble, the standard quantum limit of $\sigma^{2}$ is given by $\sigma_{\mathrm{sql}}^{2} = p\left(1-p\right)/N_{\mathrm{tot}}$, where $p$ is the probability of finding an atom in the excited state, and can be estimated as $p = \langle \hat{S}^{z} \rangle / N_{\mathrm{tot}} +1/2$.

We perform measurements under three conditions in order to probe the time evolution of the quantum noise distribution: $\tau_{\mathrm{dark}}=\{20\; \mathrm{ms},40\; \mathrm{ms}, 60\; \mathrm{ms} \}$. For these dark times, we used $\pi$-pulse times of $T_{R} =  \{10\; \mathrm{ms},20\; \mathrm{ms}, 20\; \mathrm{ms} \}$, respectively.  We find that utilizing long $\pi$~pulses reduces the sensitivity to spurious high-frequency components of laser noise. As summarized in Fig.~\ref{Fig5}, for $N_{\textrm{tot}}=1\times 10^{3}$, the quantum noise contribution to the spin noise is comparable to that of the laser noise. However, with $N_{\textrm{tot}}=4\times 10^{3}$, the laser noise is responsible for a larger fraction of the noise in repeated measurements of $\langle \hat{S}^{z}\rangle$.
There are qualitative differences between the low and high atom number cases; for example, for $N_{\textrm{tot}}=4\times 10^{3}$ with $\tau_{\mathrm{dark}}=20$~ms and 40~ms, we observe a phase shift for the minimum of the spin noise. We have verified that in the limit of small $\tau_{\mathrm{dark}}$, the phase shift is no longer prevalent.
To compare the predictions of the full many-body master equation with the experiment, we add the effect of laser noise in quadrature with the calculated spin quantum noise. We additionally find it necessary to fully treat the effects of interactions during the laser pulses. As shown in Fig.~\ref{Fig5}, the theory predicts a direction and magnitude of the phase shift of the noise minimum that agree with the experimental observations. The theory also predicts significant spin noise for rotations near $\pm 90^{\circ}$ due to an effect analogous to anti-squeezing. Despite the high sensitivity to laser noise at this rotation quadrature, the measurements of the total noise are consistent with theory.

\section{Conclusion and outlook}

The use of an intricate competition between coherent interactions and dissipation as a means to enhance and engineer many-body correlations is an emerging theme in modern quantum science\cite{Diehl2010,Barreiro2011,Daley2009}. We expect that the capabilities demonstrated here will develop into important tools to explore the rich interplay between many-body effects and dissipation in alkaline-earth atoms confined in optical lattices, as well as in other systems such as polar molecules\cite{Ni:2008wo, deMiranda:2011gd, Chotia2012, Gorshkov2011, Kaden2012}. Furthermore, as a matter of practical importance, these techniques could be used to generate interaction-induced spin squeezing in an optical lattice clock\cite{Foss-Feig2012}. Although the investigation described here is restricted to nuclear-spin-polarized gases, exploration of similar many-body effects in a clock making use of additional nuclear spin degrees of freedom is a promising first step towards the investigation of SU(\emph{N}) orbital magnetic models in ultracold atoms.

\break

\break

\clearpage

\break
\section*{Supplementary material}

\subsection*{Experimental procedures}
\subsubsection*{Atom preparation}
We load fermonic ${}^{87}$Sr into a one-dimensional vertically oriented optical lattice via two-stage laser cooling, directly producing lattice-trapped samples with temperatures of 3--5 \micro K. The optical lattice is maintained near the magic wavelength\cite{Takamoto:2005ef,Ye:2008fs} for the $^{87}$Sr $^{1}$S$_{0} \rightarrow {}^{3}$P$_{0}$ clock transition. After further cooling and optical pumping into  ${}^{1}$S$_{0}$ $\ket{F=9/2,\: m_{F} = +9/2}$,
we obtain $5\times 10^3$ nuclear spin polarized atoms concentrated mainly in a central region of approximately 100 disk-shaped lattice sites with typical trap frequencies as described in the text. We determine the longitudinal temperature, $T_{Z}$, via sideband spectroscopy and the radial temperature,  $T_{R}$, via Doppler spectroscopy \cite{Blatt:2009iq, Swallows:2011er}. With typical sample temperatures of $T_{Z}  =1-2$~\micro K, $T_{R}  =2-4$~\micro K, we find that for our trapping conditions the average density is $5(2)\times 10^{11}$~cm$^{-3}$ for a sample consisting of $5\times 10^3$ atoms. With the sample prepared as outlined above, we perform high-resolution spectroscopy on the ${}^{1}$S$_{0} \; \ket{F=9/2,\: m_{F} = +9/2}   \rightarrow {}^{3}$P$_{0} \; \ket{F=9/2,\: m_{F} = +9/2}$ clock transition.

\subsubsection*{Contrast measurement}
As described in the text, we measure Ramsey fringe contrast as a function of initial pulse area.
For a first pulse of area $\theta_{1}$, we allow the system to evolve for time $\tau$. We then apply a final pulse of area $\pi/2$ and measure the resulting excitation fraction as a function of the optical phase of the second pulse relative to the first pulse. For $\tau \gtrsim 100$~ms, there is a significant additional random phase added due to the frequency fluctuations of the ultrastable clock laser. A given excitation fraction ($p_{i}$) measurement will yield  $p_{i}=\mathcal{C} \sin^{2} \left(\Delta \phi _{i}\right)$, where $\mathcal{C}$ is the contrast and $\Delta \phi_{i}$ is the $i$th realization of the both deterministically and randomly varied phase. By analyzing  $\mathrm{Var}\left(p\right) = \mathcal{C}^{2}$/8, and assuming a uniform distribution of $\Delta \phi_{i}$, we obtain the contrast in a way that is insensitive to the laser noise.

\subsubsection*{Data analysis for spin noise measurement}

We perform quadrature-dependent spin noise measurements as described in the text at a given target atom number $N_{\mathrm{tot}} = 1\times10^{3}$ or $N_{\mathrm{tot}} =4\times10^{3}$. During the course of these measurements, we typically observe slow, systematic fluctuations of the atom number on the order of $\pm 10$\% as we operate the experiment and measure spin noise over the course of $\sim$10~hours. In most instances, these fluctuations are negligible due to the normalization techniques we employ. However, the atomic spin noise depends directly on the atom number, and a slowly varying atom number could result in unintended systematic biases. Specifically, spin noise for the coherent spin state typically considered in optical clocks scales as $1/\sqrt{N_{\mathrm{tot}}}$. Thus, the deviations in atom number can cause variations on the order of $\pm 5$\% in the measured spin noise. Ideally, these fluctuations are randomly distributed and should not result in interpretation as a false-positive for non-trivial spin-noise correlations. In the unlikely possibility that these fluctuations were correlated with a specific measurement quadrature, they could cause a spurious phase shift in the spin noise minimum. We thus take care to analyze the data in a way that is immune to this potential bias.  In this supplement, we describe our process for removing the variability in the spin noise data due to a slowly fluctuating total atom number. In this way, we verify that the phase shifts measured in Fig.~5 are not manifestations of the ``trivial'' case, where the spin noise is described by a coherent spin state. Rather, we verify that the phase shift of the spin noise minimum is a direct result of the many-body interaction as described in the Article and Methods.  

A given measurement of $\langle \hat{S}^{z}_{\mathrm{tot}} \rangle /N_{\mathrm{tot}}$ is accomplished by independently measuring $N_{e(g)}$, the number of atoms in the excited (ground) state after a single Ramsey experimental sequence, using standard electron shelving techniques. We determine its $i$th value, $\langle \hat{S}^{z}_{\mathrm{tot}} \rangle_{i} /N^{i}_{\mathrm{tot}}$, by measuring the $i$th value of  $N_{e(g)}$ (which we denote as $N^{i}_{e(g)}$) and obtain
\begin{equation}
\langle \hat{S}^{z}_{\mathrm{tot}} \rangle_{i} /N^{i}_{\mathrm{tot}} = \frac{N^{i}_{e}}{N^{i}_{e}+N^{i}_{g}} -1/2. 
\end{equation} 
From the $j$th set of  measurements of $\langle \hat{S}^{z}_{\mathrm{tot}} \rangle$, denoted $\{\langle \hat{S}^{z}_{\mathrm{tot}} \rangle_{1}, \hdots, \langle \hat{S}^{z}_{\mathrm{tot}} \rangle_{i}, \hdots \langle \hat{S}^{z}_{\mathrm{tot}} \rangle_{n_{j}} \}_{j}$, we estimate $\sigma^{2}_{j} \equiv  \langle \left(S_{\mathrm{tot}}^{z} \right)^{2} \rangle/N_{\mathrm{tot}}^{2}  - \langle \hat{S}_{\mathrm{tot}}^{z} \rangle ^{2}/N_{\mathrm{tot}}^{2}$ using a pair variance, such that
\begin{equation}
\sigma_{j}^{2} = \frac{1}{2\left(n_j-1\right)} \sum_{i=1}^{i=n_{j}} \left(\langle \hat{S}^{z}_{\mathrm{tot}} \rangle_{i+1}-\langle \hat{S}^{z}_{\mathrm{tot}} \rangle_{i}\right)^{2}. 
\end{equation}
For white noise, the pair variance is a good estimator for the standard deviation \cite{Rutman:1978cd}, while remaining insensitive to noise processes that only manifest themselves on long time scales. The number of measurements in a set, $n_{j}$, was typically $n_{j}\simeq 80$. For a given measurement quadrature, we average the results of many such measurement sets to produce one experimental data point (\textit{i.e.}, a data point in Fig.~5).

In order to maintain insensitivity to slow fluctuations in atom number between sets $j$ and $j'$, we consider the standard expression for quantum noise for the case of a coherent spin state, $\sigma_{\mathrm{sql}}$, which is expected in the absence of many-body interactions. The explicit goal is to remove any mechanism by which the trivial case---where the spin noise is described by $\sigma_{\mathrm{sql}}$---can mimic the many-body effect we predict from the theory. We calculate the $j$th value of $\sigma_{\mathrm{sql}}$ as 
\begin{equation}
\left(\sigma^{j}_{\mathrm{sql}}\right)^2 = p_{j} \left(1-p_{j}\right)/N_{\mathrm{tot}}^{j} ,
\end{equation}
where $p_{j} = \mathrm{Mean}\left[\{N_{e}^1/ \left(N_{e}^1 +N_{g}^1\right), \hdots, N_{e}^i/ \left(N_{e}^i +N_{g}^i\right), \hdots, N_{e}^{n_{j}}/ \left(N_{e}^{n_{j}} +N_{g}^{n_{j}} \right)   \}_{j}\right]$. We additionally consider a technical noise term, which represents the effect of intrinsic technical detection noise, given by $\Delta s_{j}$. This noise is characterized by a separate measurement. The detection noise accounts for 10\% of the observed noise at typical low atom numbers, while at high atom number it is only $\sim1$\% of the observed noise, and is therefore negligible. It is quadrature-independent in all cases.  From the $\sigma_{j}^{2}$, we subtract the atom-number-dependent $\left(\sigma^{j}_{\mathrm{sql}}\right)^2$ such that
\begin{equation}
\tilde{\sigma_{j}}^{2} = \sigma_{j}^{2} - \left(\sigma^{j}_{\mathrm{sql}}\right)^2 - \Delta s_{j}^{2}.
\end{equation}
Here, $\tilde{\sigma_{j}}^{2}$ represents only the effects of non-trivial spin noise and laser noise. 

The many-body theory for a given measurement condition is calculated at fixed atom number. To facilitate comparison with the many body theory, we add a noise term back to $\tilde{\sigma_{j}}^{2}$ that corresponds to $\sigma^{2}_{\mathrm{sql}}$ for the mean atom number over the entire data set, $\overline{\sigma^{2}_{\mathrm{sql}}}$. We emphasize that $\overline{\sigma^{2}_{\mathrm{sql}}}$ is a constant number, with no quadrature dependence. The many-body theoretical prediction is calculated based upon the same mean atom number used to calculate $\overline{\sigma^{2}_{\mathrm{sql}}}$. Ultimately, the net effect of this process is to remove the variability due to slow fluctuations in atom number, but to retain the part of the noise that departs from $\sigma_{\mathrm{sql}}$ due to both laser noise and many-body effects. As discussed in the text, we observe a phase shift of the minimum of the phase noise that is consistent with the many-body theory and indicative of correlated spin noise of the atom ensemble.

%
%
%

\subsection*{Derivation of the spin Hamiltonian}
The many-body Hamiltonian describing a nuclear spin-polarized ensemble of fermionic atoms  with two accessible electronic states, $g$ and $e$, which experience the same  external potential $V_{ext}(\mathbf{R})$, can be expressed as  \begin{eqnarray}
&&\hat H = \sum_{\alpha }  \! \int  \!\! d^3  \mathbf{R} \hat \Psi^\dagger_{\alpha }(\mathbf{R})   \left(- \frac{\hbar^2}{2 m} \nabla^2 + V_{ext}(\mathbf{R})\right) \hat \Psi_{\alpha }(\mathbf{R})   +\frac{4 \pi \hbar^2 a_{eg}^-}{m}   \! \int  \!\! d^3 \mathbf{R} \hat \Psi^\dagger_{e }(\mathbf{R}) \hat \Psi_{e }(\mathbf{R}) \hat \Psi^\dagger_{g }(\mathbf{R}) \hat \Psi_{g }(\mathbf{R}) \nonumber \\&&
+\frac{ 3 \pi \hbar^2}{m} \sum_{\alpha,\beta} b_{\alpha \beta}^3\! \int  \!\! d^3 \mathbf{R} \Bigg[ \Big(\vec{\nabla} \hat \Psi_{\alpha }^\dagger\Big)  \hat \Psi_{\beta }^\dagger -\hat \Psi_{\alpha }^\dagger \Big(\vec{\nabla} \hat \Psi_{\beta }^\dagger \Big) \Bigg]\cdot \Bigg[ \hat \Psi_{\beta } \Big(\vec{\nabla} \hat \Psi_{\alpha }\Big)-\Big(\vec{\nabla} \hat \Psi_{\beta }\Big) \hat \Psi_{\alpha } \Bigg] . \label{ham0}
\end{eqnarray} Here $\hat \Psi_{\alpha }(\mathbf{R})$ is a fermionic field operator at position $\mathbf{R}$ for atoms with mass $m$  in electronic state  $\alpha = g,e$. We have included only $s$-wave and $p$-wave channels, an assumption valid at $\mu$K temperatures. Since polarized fermions are in a symmetric nuclear-spin state, their $s$-wave interactions are characterized by only one scattering length $a_{eg}^-$,  describing collisions between two atoms in the antisymmetric electronic state, $\frac{1}{\sqrt{2}}(|ge\rangle-|eg\rangle)$. The $p$-wave interactions can have three different scattering volumes $b_{gg}^3$, $b_{ee}^3$, and $b_{eg}^3$, associated with the three possible electronic symmetric states (~$|gg\rangle$, $|ee\rangle$, and $\frac{1}{\sqrt{2}}(|ge\rangle+|eg\rangle)$, respectively. Note that here we have assumed no external laser field and have ignored the optical energy splitting  since the number of atoms in the excited state remains fixed  in the absence of driving terms.

In the experiment, $V_{ext}(\mathbf{R})$ is a deep 1D lattice along $Z$, which creates an array of two-dimensional discs and induces a weak harmonic radial (transverse) confinement
with an angular   frequency $\omega_R = 2 \pi \nu_R$. The lattice confines the atoms to the lowest axial vibrational mode.
We expand the field operator in a harmonic oscillator basis,
 $\hat{\Psi}_{\alpha}(\mathbf{R})=\phi^Z_0(Z) \sum_{\bf{ n}} \hat{c}_{\alpha \bf{n}} \phi_{n_X} (X) \phi_{ n_Y} (Y) $, where
$\phi_0^{Z}$ and $\phi_n$ are, respectively,  the longitudinal and the transverse  harmonic oscillator eigenmodes and $\hat{c}^\dagger_{\alpha  \bf{n}}$ creates a fermion in mode  ${\bf {n}}=(n_X,n_Y)$  and  electronic state $\alpha$.  In this basis, $\hat H$  can be rewritten as\cite{Swallows:2011er,Lemke:2011ur} 
\begin{eqnarray}
\hat H&=&\sum_{\alpha,\bf {n}} E_{\bf {n}} \hat{n}_{\alpha \bf {n}}  + \sum_{\alpha,\beta,\bf {n},\bf {n'},\bf {n''},\bf {n'''}
} \frac{\hbar}{4}\Bigg ((1-\delta_{\alpha,\beta} ) u S_{\bf {n} \bf {n'}\bf {n''} \bf {n'''}}+ v^{\alpha, \beta}
P_{\bf {n} \bf {n'}\bf {n''} \bf {n'''}} ^{(2D)}\Bigg)\hat{c}^\dag_{\alpha \bf {n}}\hat{c}^\dag_{\beta \bf {n'}}\hat{c}_{\beta \bf {n''}}\hat{c}_{\alpha \bf {n'''}}, \notag \\
u&=& \sqrt{\omega_{Z} \omega_{R}} \frac{ a_{eg}^-}{a_{ho}^R},\quad \quad
v^{\alpha, \beta}= \sqrt{\omega_{Z} \omega_{R}} \frac{ b_{\alpha,\beta}^3}{{a_{ho}^R}^3}
\end{eqnarray} $\delta_{\alpha,\beta}$ is a Kronecker delta function. Here, $a_{ho}^R=\sqrt{\hbar/(m\omega_R)}$ is the radial harmonic oscillator length, and  $S_{\bf {n} \bf {n'}\bf {n''} \bf {n'''}}$ and $P_{\bf {n} \bf {n'}\bf {n''} \bf {n'''}}$ characterize $s$- and $p$-wave matrix elements  respectively which depend on the harmonic oscillator modes and satisfy   $S_{\bf {n} \bf {n'}\bf {n''} \bf {n'''}} = S_{\bf {n} \bf {n'}\bf {n'''} \bf {n''}} = S_{\bf {n'} \bf {n}\bf {n''} \bf {n'''}} = S_{\bf {n'} \bf {n}\bf {n'''} \bf {n''}}$ and  $P_{\bf {n} \bf {n'}\bf {n''} \bf {n'''}} = -P_{\bf {n} \bf {n'}\bf {n'''} \bf {n''}} = -P_{\bf {n'} \bf {n}\bf {n''} \bf {n'''}} = P_{\bf {n'} \bf {n}\bf {n'''} \bf {n''}}$. $E_{{\bf n}}$ are single-particle energies in the trap.

Under typical operating conditions, $\nu_R\sim 450$~Hz, the interaction energy is about two orders of magnitude weaker than the single-particle energy. Thus, at the leading order, only collision events that conserve the total single-particle energy need to be considered. For a slightly anharmonic spectrum (the anharmonicity is weak but can be at the level of the interaction energy), with only one exception\footnote{The only exception are the processes $(n_{X_1},n_{Y_1})$ and $(n_{X_2},n_{Y_2})\to (n_{X_2},n_{Y_1})$ and $(n_{X_1},n_{Y_2})$.}, those processes conserve the number of particles per mode. In this case, for an initial state with at most one atom per mode ($|g\rangle$-polarized state),  it is possible to reduce  $\hat H$  to a spin-$1/2$ model.
\begin{eqnarray}
&&\hat H^{S}/\hbar = \sum_{j\neq j'}^N[ J^\perp_{{\bf n}_j ,{\bf n}_{j'}} (\vec{{ S}}_{{\bf n}_j}\cdot \vec{{ S}}_{ {\bf n}_{j'}}) +\chi_{{\bf n}_j ,{\bf n}_{j'}}{ \hat S}^z_{{\bf n}_j}{\hat  S}^z_{ {\bf n}_{j'}}
+ \frac{C_{{\bf n}_j ,{\bf n}_{j'}}}{2} ({\hat  S}^z_{{\bf n}_j}+{\hat  S}^z_{{\bf n}_j'})]. \label{many}
\end{eqnarray}
Here  $\vec{{S}}_{{\bf n}_j} = \frac{1}{2}\sum_{\alpha,\beta}\hat{c}^\dag_{\alpha {\bf n}_j}\vec{\sigma}_{\alpha \beta}\hat{c}_{\beta {\bf n}_j}$, with $\sigma^{x,y,z}_{\alpha \beta}$ Pauli matrices.  Constant terms have been dropped. \begin{eqnarray}
 J^\perp_{{\bf n}_j ,{\bf n}_{j'}}=\frac{V^{eg}_{{\bf n}_j ,{\bf n}_{j'}}-U^{eg}_{{\bf n}_j ,{\bf n}_{j'}}}{2},\quad
\chi_{{\bf n}_j ,{\bf n}_{j'}}=\frac{V^{ee}_{{\bf n}_j ,{\bf n}_{j'}}+ V^{gg}_{{\bf n}_j ,{\bf n}_{j'}}-2V^{eg}_{{\bf n}_j ,{\bf n}_{j'}}}{2},\quad
C_{{\bf n}_j ,{\bf n}_{j'}}= \frac{(V^{ee}_{{\bf n}_j ,{\bf n}_{j'}}-V^{gg}_{{\bf n}_j ,{\bf n}_{j'}})}{2}
\end{eqnarray} The quantities
 \begin{eqnarray}
  V^{\alpha\beta}_{{\bf n}_j ,{\bf n}_{j'}}=v^{\alpha, \beta}P_{{\bf n}_j ,{\bf n}_{j'},{\bf n}_{j'} ,{\bf n}_{j}}\equiv v^{\alpha, \beta} P_{{\bf n}_j ,{\bf n}_{j'}},  \quad  U^{eg}_{{\bf n}_j ,{\bf n}_{j'}}=u S_{{\bf n}_j ,{\bf n}_{j'},{\bf n}_{j'} ,{\bf n}_{j}}\equiv u S_{{\bf n}_j ,{\bf n}_{j'}}, \label{qua}
 \end{eqnarray} encapsulate the temperature dependence of the interactions. A further  simplification of Eqn.~\ref{many} can be made thanks to
 the following considerations: (i) The atoms  start in the totally symmetric manifold with $S=N/2$ since at   time $t=0$ all the atoms are polarized in the $g$ state; (ii) $P$-wave interactions are suppressed by the centrifugal barrier, and thus, at $\mu$K temperatures,   $ U$ should in general dominate over $ V$ (although it must be said that  the actual values of the $s$ and $p$ scattering parameters are not known); (iii) The $p$-wave matrix elements, $P_{{\bf n}_j ,{\bf n}_{j'}}$, are  functions which weakly dependent on ${{\bf n}_j ,{\bf n}_{j'}}$. All of these  considerations generate an energy gap that prevents transitions between the $S=N/2$ and $S=N/2-1$ sectors. Consequently, to a very good approximation the dynamics can be projected into the $S=N/2$ manifold with an effective Hamiltonian given by Eqn.~1 (which also contains single-particle terms arising from the optical driving field). $\chi=\frac{\sum_{j\neq j'} \chi_{n_j,n_{j'}}}{N(N-1)}$
and  $C=\frac{\sum_{j\neq j'} C_{n_j,n_{j'}}}{N(N-1)}$ are mode-averaged quantities. We  have omitted the term $\vec{S}\cdot\vec{S}$, which is a constant of motion.

\subsection*{Decoherence and two-body losses -- Exact treatment}

The Hamiltonain formulation described above is valid only for a closed system.  To account for losses due to inelastic $e$-$e$ collisions,
one needs to use instead a master equation:
\begin{equation}
\hbar \frac{d}{dt} {\hat \rho}= -{\rm i}[ {\hat H} , {\hat \rho} ] +  {\mathcal L}{\hat\rho}. \label{master}
\end{equation} Here  $\hat \rho$ is the  reduced density matrix operator of the many-body system. $\hat H$ is the  Hamiltonian given by Eqn.1,  and $ {\mathcal L}$ is a Liouvillian  that accounts for inelastic processes. Considering $p$-wave $e$-$e$ losses and under the same assumption of frozen motional degrees of freedom, $ {\mathcal L}$ is given by
    \begin{eqnarray} \mathcal L&=&\sum_{j\neq j'}  \frac{\hbar}{2}\Gamma_{{\bf n}_j,{\bf n}_{j'}}\big [2\hat{A}_{{\bf n}_j,{\bf n}_{j'}} {\hat \rho} (\hat{A}_{{\bf n}_j,{\bf n}_{j'}} )^\dagger-(\hat{A}_{{\bf n}_j,{\bf n}_{j'}} )^\dagger \hat{A}_{{\bf n}_j,{\bf n}_{j'}}  {\hat \rho} - {\hat \rho}
(\hat{A}_{{\bf n}_j,{\bf n}_{j'}} )^\dagger \hat{A}_{{\bf n}_j,{\bf n}_{j'}} \big ]. \label{master}
\end{eqnarray} Here the jump operators are $\hat{A}_{{\bf n}_j,{\bf n}_{j'}}=\hat{c}_{e {\bf n}_j}\hat{c}_{e {\bf n}_{j'}}$ and $\Gamma_{{\bf n}_j,{\bf n}_{j'}}=\gamma^{ee} P_{{\bf n}_j {\bf n}_{j'} }$. The expression for $\gamma$ is identical to $v^{e,e}$ up to the replacement of the $p$-wave elastic scattering volume by the inelastic one.

An important note is that there is no coherence between sectors of different atom numbers and the master equation can be solved in a ``block-diagonal'' way (see Fig.~1). Specifically, if the system starts with $N$ particles, we need to solve a series of differential equations for each of the subspaces with cascading atom numbers: first for $N$ particles, next for $N-2$ particles, then for  $N-4$ particles, etc.   There are ${N \choose n}\equiv \frac{N!}{n!(N-n)!}$ different sectors with $ N-n$ particles, $n=0,2\dots N$. For mode independent interaction parameters $\Gamma_{{\bf n}_j,{\bf n}_{j'}}\to \Gamma$. In each of the $\mathcal{N}=N-n$ sectors, the dynamics is restricted to the collective Dicke states $|S=\mathcal{N}/2,M_\mathcal{N}\rangle\equiv|M_\mathcal{N}\rangle$. Moreover, each of the  ${N \choose n}$ sectors behaves identically.

Let  $\rho_\mathcal{N}$ be the density matrix for a single sector of $\mathcal{N}$ particles. Furthermore, let us assume that particles in $\rho_\mathcal{N}$ are numbered from 1 to $\mathcal{N}$ in such a way that atoms $\mathcal{N}+2$ and $\mathcal{N}+1$ are the ones that decay as one goes from $\rho_{\mathcal{N}+2}$ to $\rho_\mathcal{N}$. The resulting equations are
{\small
\begin{eqnarray}
\frac{d}{dt} \rho_{\mathcal{N} } = - \frac{i}{\hbar} [H_{\mathcal{N} },\rho_{\mathcal{N} }] - \frac{ \Gamma}{2} \sum^{\mathcal{N} }_{i < j} (\hat n_{ei} \hat n_{ej} \rho_{\mathcal{N} } + \rho_{\mathcal{N} } \hat n_{e i} \hat n_{e j}) + \Gamma {N-\mathcal{N}  \choose 2}(\hat  c_{e,\mathcal{N} +2} \hat c_{e,\mathcal{N} +1} \rho_{\mathcal{N} +2} \hat c^\dagger_{e,\mathcal{N} +1}\hat  c^\dagger_{e, \mathcal{N} +2}),
\end{eqnarray}}for $0\leq\mathcal{N}\leq N$. In terms of spin operators,
{\small\begin{eqnarray}
&&\sum_{i < j}^\mathcal{N} \hat n_{e i} \hat n_{e j}
= \frac{\mathcal{N} (\mathcal{N}-2)}{8} + \frac{\mathcal{N}-1}{2} \hat S^z + \frac{1}{2}(\hat S^z)^2, \end{eqnarray}
\begin{eqnarray}
\langle M_\mathcal{N} | \hat c_{e,\mathcal{N} +2} \hat c_{e,\mathcal{N} +1} \rho_{\mathcal{N} +2} \hat c^\dagger_{e,\mathcal{N} +2} \hat c^\dagger_{e,\mathcal{N} +1}|M_\mathcal{N} '\rangle =
\langle  M_{\mathcal{N}+2}+2|\rho_{\mathcal{N}+2}| M_\mathcal{N+2}'+2\rangle \sqrt{\frac{{\mathcal{N}\choose M_\mathcal{N}+\mathcal{N}/2} {\mathcal{N}\choose M'_\mathcal{N}+\mathcal{N}/2}}{{\mathcal{N}+2 \choose M_\mathcal{N} +\mathcal{N}/2+2} {\mathcal{N}+2 \choose M'_\mathcal{N} +\mathcal{N}/2+2}}}.
\end{eqnarray}} For $N \leq 50$,  the above equations can be efficiently solved numerically. \footnote{ At the highest operating densities a cut-off of 50 atoms per lattice site  is sufficient.}

\subsection*{Decoherence and two-body losses -- Mean-field-treatment}

A simple and illuminating way to perform a mean-field treatment is to use the Schwinger boson representation that maps spin operators to two-mode bosons \cite{Auerbach1994}. It represents the spin operators as
\begin{eqnarray}
2{\hat S}^z=\hat{\Psi}_1^\dagger \hat{\Psi}_1-\hat{\Psi}_0^\dagger \hat{\Psi}_0 \quad \quad
{\hat S}^+=\hat{\Psi}_1^\dagger \hat{\Psi}_0 \quad \quad {\hat S}^-&=&\hat{\Psi}_0^\dagger \hat{\Psi}_1 ,
\end{eqnarray}with $\hat{\Psi}_j$ a bosonic annihilation  operator of mode $j=0,1$. In terms of Schwinger bosons, the collective spin model maps exactly to the  Hamiltonian that describes a double-well bosonic system (by identifying the 0 and 1 states with the left and right wells)  with a tunneling matrix element $\Omega$,  a bias between the  wells $\delta$,  on-site interactions $V^{gg},$ $V^{ee}$ in the left and right wells, respectively, and nearest-neighbor interactions $V^{eg}$. The mean-field approximation, which gives rise to the Gross-Pitaevskii equation, replaces the field operators $\hat{\Psi}_{1,0}$ with complex numbers, $\hat{\Psi}_{1,0}\to \Phi_{0,1}$.  To deal with the lossy dynamics, it is necessary to express the equations of motion in terms of the density matrix, $\rho_{ij}\equiv{\Phi}_{i}^*{\Phi}_{j}$. The resulting equations are
\begin{eqnarray}
 \frac{\partial}{\partial t} {\rho}_{00}&=&  -i\frac{\Omega}{2} ({\rho}_{10}-{\rho}_{01}),\\
 \frac{\partial}{\partial t} {\rho}_{11}&=& i\frac{\Omega}{2} ({\rho}_{10}-{\rho}_{01})- \Gamma{\rho}_{11}^2,\\
 \frac{\partial}{\partial t} {\rho}_{10}&=& i\frac{\Omega}{2} ({\rho}_{11}-{\rho}_{00})+i \left [-\delta+C({\rho}_{11}+{\rho}_{00}) + \chi ({\rho}_{11}-{\rho}_{00})+ i  \frac{\Gamma {\rho}_{11}}{2}\right ] {\rho}_{10},
\label{gpe3}
   \end{eqnarray} where ${\rho}_{01}={\rho}_{10}^*$. The mean-field treatment is expected to be valid in the regime $\sqrt{N}\sin[\theta_1]\chi t\ll 1$.

\subsection*{Higher Order Corrections}

The spin Hamiltonian neglects collision processes that do not preserve single-particle energy. However, collisions populating off-resonant states can still take place virtually and will introduce corrections to the spin model. To account for those, we split  the Hilbert space into the resonant, $\Sigma$, and non-resonant, $\Upsilon$,  manifolds respectively, spanned by the states
\begin{eqnarray}
|\Phi^\Sigma_{\vec{\bf\sigma}_{\vec{n}}} \rangle&=&|\sigma_{{\mathbf n}_1},\sigma_{{\mathbf n}_2},\dots,\sigma_{{\mathbf n}_N}\rangle,\quad E_0^{tot}\equiv\sum_{j=1}^N E_{{\bf n}_j},\\
|\Psi^\Upsilon _{\vec{ \bf \sigma}_{\vec{k}}}\rangle&=&|\sigma_{{\mathbf k}_1},\sigma_{{\mathbf k}_2},\dots,\sigma_{{\mathbf k}_N}\rangle, \quad E_{\bf {\vec k}}^{tot}\equiv\sum_{j=1}^N E_{{\bf k}_j}\neq E_0^{tot}.
\end{eqnarray} These states are  written in the occupation basis and $\sigma \in  \{g,e\}$. In $|\sigma_{{\mathbf k}_j}\rangle$, the same mode ${\mathbf k}_j$ can also be occupied by both an $e$ and a $g$ atom simultaneously.

By projecting the many-body Hamiltonian on $\Sigma$ using the corresponding projection operator ${\mathcal P}_\Sigma$ one can obtain an  effective Hamiltonian:
\begin{eqnarray}
\hat H^{\rm eff}={\mathcal P}_\Sigma {\hat H}{\mathcal P}_\Sigma={ \hat H}^{S}+ { \hat  H}^{S_2}.
\end{eqnarray} Here, $\hat H^{S}$ is the spin model given by Eqn.~\ref{many}. $ {\hat H}^{S_2}$ can be obtained via the second order perturbation theory as follows:
\begin{eqnarray}
\langle \Phi^\Sigma_{\vec{\sigma}_{\vec{n}}} |{\hat H}^{S_2}|\Phi^\Sigma_{\vec{\sigma'}_{\vec{n}}} \rangle&=&-\sum_{{\bf \sigma}_{\vec{k}}}\frac{\langle \Phi^\Sigma_{\vec{\sigma}_{\vec{n}}}|{\hat H}|\Psi^\Upsilon _{\vec{ \bf \sigma}_{\vec{k}}}\rangle\langle \Psi^\Upsilon _{\vec{ \bf \sigma}_{\vec{k}}}|{\hat H}|\Phi^\Sigma_{\vec{\sigma'}_{\vec{n}}}\rangle}{ E_{\bf {\vec k}}^{tot}-E_0^{tot}},
\end{eqnarray}

 When ${\hat H}^{S_2}$ is written in terms of spin operators and is projected into the collective Dicke manifold, it gives rise to terms proportional to $\hat {S}^z$,$(\hat S^z)^2$ and $(\hat S^z)^3$. We note that terms proportional to $(\hat S^z)^4$ vanish for $p$-wave interactions. Among those corrections, the first two can be absorbed into the spin Hamiltonian given by Eqn.~1 and only the third one gives rise to additional corrections. The corrections in the Ramsey contrast decay arising from the cubic term are shown in Fig. S1.

\begin{figure}[h!]
\centering
\includegraphics[width=0.65\columnwidth]{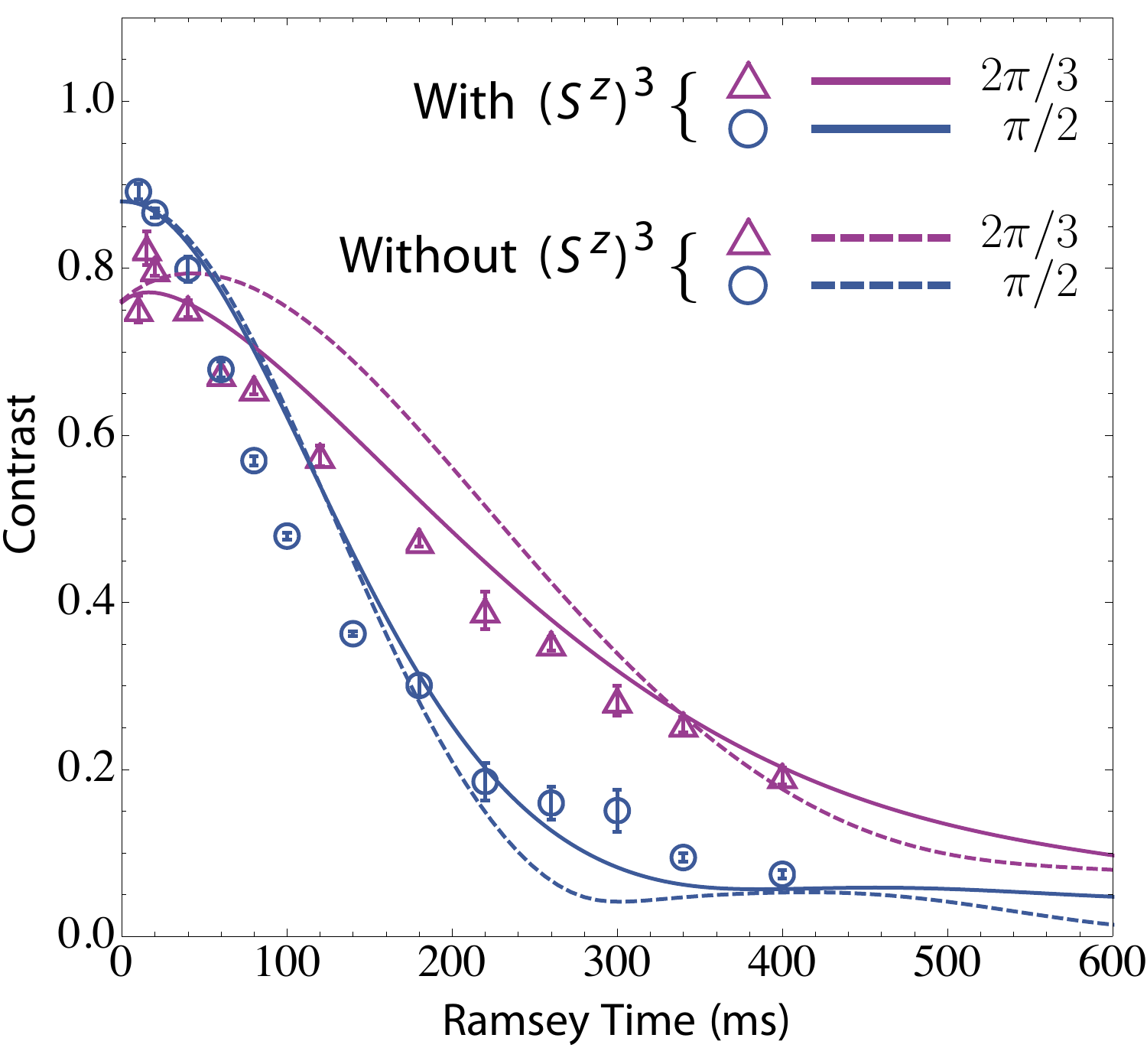}
\caption{\textbf{Ramsey fringe contrast decay vs initial pulse area.}   The solid lines are the many-body fits using terms of order $\left(\hat{S}^z\right)^{3}$, and the dashed lines represent fits without these terms. These data are the same as in Fig.~4a of the main text, and were taken with $\nu_{Z}=80$~kHz and $N_{\mathrm{tot}} = 4 \times 10^{3}$. Pulse areas are shown in the legend.}  \label{decayprob}
\end{figure}

\bibliographystyle{Science}
\bibliography{ManyBib}

\begin{scilastnote}
\item We thank S. Blatt, J. Thomsen, W. Zhang, T. Nicholson, J. Williams, B. Bloom, S. Campbell, and A. Daley for discussions and technical help. The work is supported by NIST, DARPA, AFOSR, NSF, and the ARO-DARPA-OLE program. M.~B. acknowledges support from NDSEG. A.~V.~G. acknowledges support from IQIM, the Lee A. DuBridge foundation, and the Gordon and Betty Moore foundation. M.~D.~S. is currently at AOSense, Sunnyvale, California 94085, USA. J.~v.-S. is currently at Tech-X Corporation, Boulder, Colorado 80303, USA. Correspondence should be addressed to M.~J.~M. (michael.j.martin@colorado.edu), A.~M.~R (arey@jilau1.colorado.edu), or J. Y. (ye@jila.colorado.edu). 
\end{scilastnote}

\end{document}